# Satellite Based IoT for Mission Critical Applications

Sudhir K. Routray
Department of Electrical Engineering
Addis Ababa Science and Technology University
Addis Ababa, Ethiopia
Email: sudhir.routray@aastu.edu.et

Abhishek Javali
Department of Telecommunication Engineering
CMR Institute of Technology
Bangalore, India
Email: abhishek.j@cmrit.ac.in

Laxmi Sharma
Department of Telecommunication Engineering
CMR Institute of Technology
Bangalore, India
Email: laxmi.sh@cmrit.ac.in

Richa Tengshe
Department of Telecommunication Engineering
CMR Institute of Technology
Bangalore, India
Email: richa.t@cmrit.ac.in

Sutapa Sarkar
Department of Electronics and Communication Engineering
CMR Institute of Technology
Bangalore, India
Email: sutapa.s@cmrit.ac.in

Aritri D. Ghosh
Department of Telecommunication Engineering
CMR Institute of Technology
Bangalore, India
Email: aritri.d@cmrit.ac.in

*Abstract*—In the recent years, world has witnessed the ubiquitous applications of Internet of things (IoT) for many different scenarios. There are several critical applications where the results are essential and the mission has to be successful at any cost. Such applications are well known as mission critical applications. These applications are really critical and deal with very serious situations such as disaster management, rescue operations and military applications. IoT can provide both accuracy and sustainability in these applications. IoT in fact, is suitable for several critical applications because it can be deployed at locations where human presence is not possible due to the dangers to human life. In such cases, collection of information can be done through IoT sensors and it can be sent directly to the processing hubs. These days we find several mission critical applications where both increased reliability and coverage have very high priorities. Hybridization of IoT and satellite networks can be a game changer in these applications. In this article, we present the general features of mission critical IoT and the motivation for connecting it with the satellite networks. Then we present the main deployment related issues of these hybrid networks. We focused on the hybridization aspects of narrowband IoT (NBIoT) with the satellite networks. Because NBIoT has the energy efficiency which can make the satellite based IoT networks sustainable in the long term.

I. INTRODUCTION

Mission critical applications are very important as they deal with very serious circumstances and their failure may result in big losses of lives and properties. In mission critical military applications, protection of borders and defense of territories are associated. Therefore, these applications have to succeed to protect the precious resources. Internet of things (IoT) is a natural technological choice for the mission critical applications. IoTs are deployed through a large number of sensors and these sensors collect the information according to the logics set in their systems. In mission critical applications, there are a lot of risks and dangers to the human lives. Thus the information collection and actuation should be carried out by the IoT sensors. Integration of satellites with the IoTs increases the coverage and reliability of these networks for mission critical applications.

IoT is the pervasive expansion of the Internet. It spreads the connectivity beyond the people and includes billions of machines, devices and sensors. In [1], the basic features of IoT, its utilities and capabilities are presented. In this survey, the enabling technologies behind IoTs and the new protocols are analyzed for its practical deployment. Now, the market for IoT has gained pace and has a strong growth trend. In [2], the market potentials of IoT in the industries have been focused. The market for IoT is going to be much larger than it is predicted now. That is due to the new initiatives for the modern industrialization where IoT has big roles in every part of the industry [2]. It seems that IoT is going to be the main player for the next industrial revolution where automation is the main aim [3]. Due to the ubiquitous presence of IoT sensors and its actuator components, energy efficiency is an important aspect of IoT. Therefore, the green initiatives in IoTs are essential for their long term sustainability. In [4] and [5], the green initiatives in IoTs are presented with the emphasis on their practical deployment. It is found that narrowband IoT (NBIoT) is one of the cost effective and sustainable forms of IoTs [4]. It consumes less energy and bandwidth than the other forms of IoTs while providing the performances comparable to the other forms. Smart world initiatives such as smart cities and smart policing need the green initiatives as the size of the IoTs for these applications is really huge [5]. In [6], the time lines for 4.5G and 5G are presented. At present, most cities in the world use 4.5G which has good compatibility with IoT systems. After 2020, several cities will have 5G which can accommodate an enhanced deployment of IoTs for various applications [6]. The changing trends of the technologies and markets directly affect the economic growth and development. In [7], the effect of consumer electronics on overall information and communication technology (ICT) market is analyzed. It shows strong worldwide growth trends for Internet based applications including the IoT. In the recent days, IoTs have several applications which simplify the modern complexities. In [8] – [11], several applications of IoTs have been discussed. IoTs provide sensing, detection, decision making and actuation through their networks. Using these basic functions many algorithms can be developed to carry on millions of functions. The main applications these days are: smart cities, smart agriculture, large scale manufacturing, policing, retail management, traffic monitoring, crowd sensing, surveillance and mission critical operations [4], [9]. Localization using IoT is very popular because it provides better accuracy than the previously used methods [10]. In addition to localization, tracking of objects and devices is also possible with an increased accuracy through the IoTs [11]. It provides efficient logistics solutions for the long distance trade and commerce. Security of the IoT networks is currently a big challenge for the long term reliability of the IoTs. Several security measures have been adopted in the IoTs. Still the security issues are not been solved in the IoT networks. In [12], quantum cryptography is proposed for IoTs. In fact, it is a promising technology for the IoTs, though its deployment is a challenging issue. In [13], end-to-end mission critical traffic flow control in 5G networks is proposed through network programming. A network slice dedicated for the mission critical traffic is the suggested solution in a softwarized network [13]. Mission critical applications through IoT are common practices these days. For the mission critical applications the IoTs are designed with specific survivable architecture [14]. The architectures are reliable under the adverse conditions. These features are appropriately explained using examples in [14]. Satellite and IoT integration is an essential technology for large scale remote monitoring and control of objects and devices [15]. It expands the coverage area and increases the reliability of remote monitoring. Low earth orbit (LEO) satellites are suitable for satellite based IoT networks. In [16], the LEO satellite based IoTs and their applications are presented. Globally connected IoT can be deployed through the satellite links. In [17], effective data collection techniques for such globally connected satellite based IoTs are presented. Large scale data transmission and IoT activities are envisioned through satellite based IoTs in [18]. Terabits of data can be transferred through high throughput satellites and it can easily support the massive IoTs on earth.

In this article, we present the basic features and necessities of a satellite based IoT for mission critical applications. We present the motivation for satellite based IoT systems. We went through the structural aspects of such satellite-IoT hybrid system and its deployment related issues. We also analyze the energy efficiency and the long term sustainability of these hybrid systems.

The remaining parts of this article are arranged in four different sections. In section II, we present the mission critical applications and their basic characteristics. In section III, we present the reasons why satellites and IoTs should be conjugated together. In section IV, we present the satellite based IoT for mission critical applications. We show some instances in which the satellite based IoT are essential for the success of the critical applications. In section V, we conclude this article with the main points.

## II. MISSION CRITICAL APPLICATIONS

As the name suggests, mission critical applications are those in which the mission's success is essential for the people or system associated. The consequences are really critical or serious if failure results in these applications. In other words, a mission critical application is required for the survival of a person or system or project or organization or business (as per the case in which it is associated). For instance, when a mission critical application does not work or fails or is interrupted, business operations are also impacted. These operations are mission critical because they are essential to a company's mission and if they fail, that can damage the entire business system. Similarly, in case of disaster management the failure of the mission may result in the loss of several lives and properties. Military proceedings too are associated with several mission critical applications. The success or failure is directly linked with the win or loss in the battles and wars. Systems such as railway operating systems, aircraft operation and control systems, online banking system, electric power systems, and many other computer systems are example of mission critical applications. Since mission critical applications are dispersed onto multiple segments, they need to sustain typical conditions. In all these complexities, IoT provides attractive solution through its real time protocols to carry out the steps involved with the minimum possible delay.

At present, regardless of all the technologies available, all existing mission critical distributed applications face performance and reliability problems. These applications may go through any of the software or hardware failure, which causes losses in time, data, money and human lives. Failure of the system can put business in fail or can put unfavorable affect on business and society. Mission-critical applications require strict data transport but also energy efficient performance. These applications are dependent upon the efficiency in on-line fault detection where data computation integrity plays a major role. There are uncountable mission critical applications which work in our surrounding to provide us various types of services indirectly supporting many businesses. Mission-critical applications and systems require that any failure that causes any types of disruption get detected and identified as soon as possible in order not to save the whole system from failure. These days, the mission critical systems deal with a lot of data and thus use several data processing and storage units. Cloud computing is also considered as a support for mission-critical applications. The enlarged reach of mission critical applications is both at used end (customer's end) as well as at business end. So mission critical computing has evolved to include mission critical "interactions" as well as mission critical "transactions". Modernized data center to support increased mission-critical service levels and application support for mission-critical interactions are the current requirements from any mission critical system. Mission critical applications are open for large number of end users and customers with higher degree of reliability.

## III. SATELLITES FOR IoT

### A. An Overview of Satellite IoT Conjugation

Satellite constellations in geostationary earth orbits (GEO) provide terabytes of capacity and mainly are used for direct to home broadcast and Internet over satellites. GEO satellite constellations are not suitable for mission critical applications due to their high latency. Due to path loss between earth and satellite and slotted nature of the geostationary orbit, terminal antennas in such systems must have large gains to close the link and higher directivity to avoid interference with adjacent satellites and systems. LEO satellite constellations when compared to GEO have advantage of low propagation delay, global coverage and small propagation loss which are the basic needs of a mission critical application. Due to lower orbit altitude (generally lower than 2000 km) the round trip time for LEO constellation is around 100 ms while that of GEO satellite constellation is 600 ms [16]. Similar to LEO some highly elliptical orbit (HEO) satellite constellations operate closer to earth compared to GEO satellites and offer the advantage of lower path loss and hence are less costlier, lighter and have low latency. Hence these are well suited for mission critical applications. However, LEO and HEO constellations have a drawback that there location keeps on changing with respect to a point on earth hence they give rise to a highly time variant communication channel and therefore need steerable antennas. An IoT terminal in a LEO or HEO network requires tailoring the waveform as well as taking care of antenna design suitable for time variant communication channel. In [17], HEO-Molniya satellites constellation is discussed. Critical IoT applications demand dependable, low latency and high-throughput connections. These requirements can be fulfilled by using high throughput satellites (HTS). HTS utilize technologies such as concentrated spot beams, higher frequency bands and frequency reuse to increase the speed and capacity of GEO/LEO/HEO satellite constellations. Additional aspects specific to mission critical application such as low SNR, time variant channel and link budget needs to be considered while tailoring the communication standard. An integrated or hybrid LEO-GEO constellation can be a potential solution for many applications. Hybrid systems may provide benefits of both such as, lower latency, flexibility, and scalability of LEO satellites. Likewise high-capacity and wide coverage of existing GEO constellations can be achieved. However such system will require an optimized routing system which routes to LEO when it requires low latency and GEO when there is a need to transmit a large volume of data at any time. For such system to be success dynamic routing must be in place.

### B. Motivation for Satellite Based IoT

Satellite communication plays a major role in providing the interconnection between the smart things which are scattered over many geographical areas. There are several instances in which the coverage area of the mission critical applications has to be large. In addition to that longevity and multicasting facilities for a large coverage area is comparatively more beneficial through the satellites. Satellites can establish a cost-effective communication between the smart objects which may be inaccessible to the rest of the world. In scenarios where huge amount of data interconnection is needed, relying on the wired and wireless internet connections is not a good option. Satellites can become an alternative and cost-effective solution. Narrow bandwidth satellites can be recycled for many IoT applications [15]. The strong need for satellite communication appears in the extreme topographies such as cliff, valley and steep slope. The chances of failure in geologic disasters are more in these networks and satellite IoT can be extremely useful in these cases giving larger network coverage. Many IoT devices are located in remote areas such as ocean, desserts and forests to cover certain applications, these IoT sensors can be given Internet connectivity by satellites. Instead of investing on base stations to build wireless communication systems for terrestrial networks one can opt for satellites since these terrestrial wireless communication infrastructures may be easily damaged by natural calamities [16]. In Fig. 1, we show the satellite based IoT in which the satellites provide coverage for IoTs and there are also inter-satellite connections for a larger scope. Multi-satellite based coverage is quite reliable for mission critical applications.

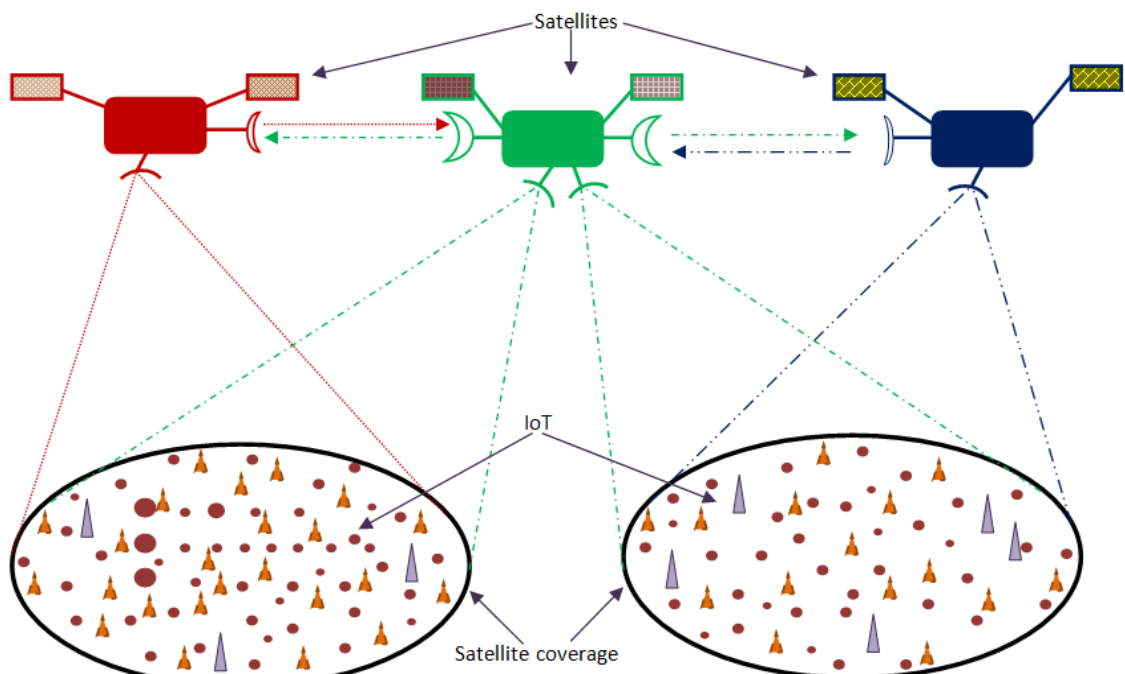

Fig. 1. Satellite based IoT network for mission critical applications. Here, we show the IoT networks which are covered by satellites and the satellites too are interconnected. The dots (in the ovals) are the sensor / actuator locations. The triangular trans-receivers are for the data transmission.

There are varieties of wireless networks which are in use to provide interconnectivity to millions of users. Some of them are ZigBee, Bluetooth, Wi-Fi and Near Filed Communication (NFC) which have become integral parts of the modern IoT ecosystem. However, the capacity of these networks is limited in terms of the number of areas covered. One of the challenges is the ability of these networks to handle multiple things and to make simultaneous interconnectivity and communication a reality. Satellite based IoT can be one of the suitable options for these concerns [17]. The future wireless communication systems are aiming at very high data rates for the users and machines to meet the high-speed demands of millions of users. The satellites can fulfill such needs of smart IoTs. Surveys done in some countries show that rural areas and tribal areas suffer from low internet connectivity. Low internet connectivity has direct consequence on the progress in the business. There is a high demand to meet the interconnectivity needs of large number of IoT devices which is projected to reach 25.6 billion units by 2020.Not only this, in corporate enterprise applications such as oil and gas platforms, where frequent monitoring is needed for security and operation, there is a strong need for high speed internet access. This has shifted the focus of researchers to the high-throughput satellites [18].

*C. Link Power budget for SIoT*
Power budget estimation for satellites is essential for each and every application. The total power required for the satellite IoT linking has to be more than the sum of the threshold power requted by the IoT sensor, power loss along the path, the noise margin and the IoT system margin. Based on the above considerations, we present equation (1) for the link power budget for satellite and IoT conjugation.

$$P_T \geq P_{IoT,Th} + P_L + P_{NM} + P_{SM} \qquad (1)$$

The symbols presented here are:
$P_T$: The total power transmitted by the satellite,
$P_{IoT,Th}$: threshold power needed by the IoT device,
$P_L$: Power loss along the channel,
$P_{NM}$: Noise margin,
$P_{SM}$: System margin.

It should be noted here that mainly low earth orbit (LEO) satellites are used for SIoT applications which are not at a constant distance from the IoT on earth. Thus the power budget terms in (1) are not constant except for $P_{IoT,Th}$.

Instead they are variables of distance and time depending on the satellites' locations with respect to the IoT networks. So, $P_T$ has to be varied according to the right hand side parameters. Alternatively, it has to be set at a higher value so that the IoT sensor terminals can always get the minimum power $P_{IoT,Th}$ or higher.

*D. Energy Efficient SIoT Networks*
Satellites and IoTs have to be bandwidth efficient. Energy and bandwidth efficient IoTs are already available for deployment. NBIoT is a standardized technology and it has been specified in LTE Release 13. According to it, NBIoT transmits at two levels: 20 dBm and 23 dBm [4]. Similarly, its receiver threshold is -64 dBm. When compared with other IoTs, NBIoTs need a very small amount of power and bandwidth. Sensors of NBIoT can be powered through small batteries or solar panels. The life spans of the batteries can be as long as ten years. Green energy harvesting is also possible for NBIoT sensors. Due to its low power requirements, NBIoT is a low power wide area network (LPWAN) technology. Therefore, NBIoT is always a first choice for low power regime. Despite that it can be used for several essential sectors such as smart cities and smart agriculture. In LTE Release 14, several new specifications for enhanced services have been mentioned. Several changes to the basic parameters of NBIoT have been proposed for multipurpose applications. These new specifications make NBIoT more versatile than its previous versions (i.e., standards of LTE Release 13). NBIoT can also be a key technology for mission critical applications. Mainly for the rural deployment it is preferred over several other forms of the IoTs. Due to its economical nature, it is also the first choice for the developing countries.

IV. APPLICATIONS OF SATELLITE BASED IOT SYSTEMS

Several applications are possible through the SIoT networks. SIoT provides larger coverage, better availability than cellular networks, better interconnection between IoTs and the Internet. Mission critical applications are more reliable through the SIoT networks due to their better availability than the cellular networks. In this section, we provide some typical applications of SIoT networks.

*A. Mission Critical Applications*
We have already explained the mission critical applications in the previous sections. During the natural disasters such as floods, cyclones, hurricanes, landslides, earthquakes and tsunamis the cellular infrastructures suffer major damages. It has been observed that these networks go out of operation for a long time. In contrast, the satellites are not at all affected by these disasters. Similarly, proper deployment of the IoT sensors can be made immune to these problems. So, the hybrid SIoT networks are preferred for these natural disasters.

*B. Location Dependent Serivices*
Location dependent services are the ones which need the exact location or position of the destination to provide their services. Cellular networks along with GPS satellites are normally used for location determination. The outcome from GPS and cellular network combination is not very accurate. However, IoT networks along with the support of the satellites can provide better outcomes [8] – [11]. Accurate location details are needed in several applications such as: military strikes, enemy object tracking and precision measurements. Therefore a GPS using satellites and IoTs is better than the other alternatives.

*C. Surface and Air Navigation Systems*

Modern navigation systems are very advanced. They need the exact motion and position related information. Due to its critical nature, these systems need real time information about their operations. This real time information can be provided through the mission critical applications of SIoT. Normally these systems need coordination between the satellites and the ground information. In SIoT, the satellites and the IoTs can do it perfectly. The dual operation of the ground based IoT sensors and the space based satellites provide much better accuracy when compared with the satellites alone.

*D. Smart Agricurture*

Agriculture is still an important economic sector. A lot of people are still associated with sector. Mainly, in the developing countries it is one of the main revenue generating sectors. Increasing the harvest is essential for this sector. However, the resources for agriculture are not available everywhere. In order to utilize the resources smartly IoTs can be utilized. Precision farming uses just the exact amount of resources and provides optimized harvest. So, it is called smart agriculture. Information for smart agriculture can be gathered through the sensors of SIoT networks. Through the SIoT sensors the farmers can monitor the water level, temperature, fertilizer concentration, humidity and several other parameters.

*E. Location Tracking*

Location tracking is a fundamental requirement in several applications. Some of the instances are: police tracks the criminals for nabbing, kids are tracked by the parents, in logistics consignments are tracked by both the senders and recipients, flights are tracked by the air traffic control, animals are tracked by their owners, pets are tracked for their safety and so on. Tracking operations need to be accurate and resource efficient. Satellite based tracking are quite effective. They provide both accuracy and efficiency.

*F. Smart Healthcare*

In very country, healthcare is a basic need. Healthcare has several aspects such as hospitalization, surgery, pre-hospitalization care, nursing, telemedicine and remote health monitoring. All these aspects can be supported by SIoT services. For instance, telemedicine and pre-hospitalization treatments can be provided through the IoTs. SIoT is preferable in these applications as their coverage is better than other IoTs and cellular networks.

SIoT application domain is too large to list all utilities here. Every year, emerging applications get added to the pool. The LEO satellites are widely used for the SIoT based applications. Even the very low earth orbit (VLEO) satellites are now deployed for SIoT operations. The net power budget for the LEO and VLEO satellites is much lower than the GEO satellites.

## V. MISSION CRITICAL APPLICATIONS OF SATELLITE-BASED IoT SYSTEMS

IoT enables a large number of services and applications which have revolutionized the way we interact with our surroundings. The ability to remotely monitor and manage objects is leading to advancement in transportation, remote healthcare, public safety, energy management, policing, retail management, home and industrial automation, wildlife tracking, naval fleet management, transparent logistics and many more. Main strengths of IoT lies in abilities to ubiquitous sensing, reliable connectivity, enhanced situational awareness, data driven decision analytics and automated response without human intervention. The number of IoT devices grows very fast and is estimated to reach several billions by 2020. It will lead towards the massive IoT which can virtually connect almost everything needed. Satellite integration with IoTs increases the capabilities of the IoTs even further. Several remote machine type communications are very much facilitated by this integration. Mission critical applications too are boosted through this new initiative. In specific mission critical scenarios, fast and accurate data reception, low-power connectivity, highly reliable and guaranteed performance are more important than high channel capacity. These cases are very much suitable for the SIoT applications. In environments where infrastructure is sparse, such as rural or remote areas, specialized infrastructure-less ad-hoc solutions are needed, which provide long-range multi-hop connectivity to remote systems. Some mission-critical applications of SIoT systems are presented in the following paragraphs.

*A. Disaster Management*

Disaster management is always mission critical due to the serious consequences they carry. Disasters such as floods, storms, cyclones, landslides, earthquakes are always associated with a lot of lives and properties. Any delay in the process can lead to very bad consequences. Accurate information on the ground situation is essential. This is possible to collect the information through the SIoTs.

*B. Deep Space Communication*

Deep space communication is always a mission critical in nature. Its operations such as separation from upper launcher stages, remote control of a decommissioning maneuver, debris mitigation, computing for control applications, spanning orbit determination, guidance, navigation control and determination of relative pose and angular rates are all critical for the success of a space mission. In this scenario, accurate data collection and its processing for the further steps is essential. Hence an SIoT system consisting of accurate sensors and actuators are required on-board as well as at the ground stations.

*C. Telemedicine and Remote Healthcare System*

In case of emergency, patients are brought to the hospital in the minimal time to provide appropriate health care. However, if the patient's location is far away from a specialized hospital then the treatments can be provided remotely through telemedicine. In this case the doctors and nurses available near the patient take the advice of the specialist doctors over a communication link. Even direct care can be provided through the sensors through SIoT. In case of transportation of patients to the hospital, appropriate treatment can be provided on the way in the ambulances through the SIoT systems. In case of both telemedicine and remote healthcare provisioning SIoT is immensely helpful.

*D. Aircraft Navigation Systems*

Aircraft is highly dependent on the air navigation system which uses various control and data aqusition for its proper operation. Satellites are placed much above the aircrafts. Thus the navigation information and even some selective control information can be provided through the SIoT to the aircrafts. Radio navigation aids helps the pilots to navigate more accurately even under low visibility. GPS is also used by pilots to provide precise location data, which includes speed, position, and track. In this case, the malfunction of navigation system would be mission critical and will cause serious consequences. Thus SIoT can play mission critical roles in air navigation process.

## VI. CONCLUSIONS

In this article, we discussed the main utilities of the satellite based IoT systems for mission critical tasks. We have gone through the basic requirements of a typical satellite based IoT system, its coverage limitations and the power needs. SIoTs are essential for several mission critical applications due to their superior performances. These systems overcome several natural difficulties of the traditional systems. For the long term sustainability, SIoTs have to be energy and bandwidth efficient. Thus resource efficient versions such as the NBIoT are preferred over other forms for the long term. Current market demand and commercial perspectives of SIoT based businesses are bright. Several companies have planned to use SIoT in businesses. Several startups companies are also showing great interest in the SIoT based services.